\documentclass[twocolumn,aps,pre,showpacs]{revtex4}
\usepackage{amssymb, amsmath}
\usepackage{graphicx}
\usepackage{epstopdf}

\newcommand{\vq}{{\bf q}}
\newcommand{\vQ}{{\bf Q}}
\newcommand{\fr}{{\mathrm{0}}}
\newcommand{\lap}{{\mathcal{L}}}
\renewcommand{\Re}{{\mathrm{Re}}}
\renewcommand{\Im}{{\mathrm{Im}}}
\newcommand{\Ical}{{\mathcal{I}}}

\newcommand{\erfc}{{\mathrm{erfc}}}
\newcommand{\ue}{{\mathrm{e}}}
\newcommand{\ud}{{\mathrm{d}}}
\newcommand{\boldnabla}{\mbox{\boldmath$\nabla$}}

\begin{document}

\title{Huygens-Fresnel-Kirchhoff construction for quantum propagators
  with application to diffraction in space and time}

\author{Arseni Goussev}

\affiliation{Max Planck Institute for the Physics of Complex Systems,
  N{\"o}thnitzer Stra{\ss}e 38, D-01187 Dresden, Germany}

\date{\today}

\begin{abstract}
  We address the phenomenon of diffraction of non-relativistic matter
  waves on openings in absorbing screens. To this end, we expand the
  full quantum propagator, connecting two points on the opposite sides
  of the screen, in terms of the free particle propagator and
  spatio-temporal properties of the opening. Our construction, based
  on the Huygens-Fresnel principle, describes the quantum phenomena of
  diffraction in space and diffraction in time, as well as the
  interplay between the two. We illustrate the method by calculating
  diffraction patterns for localized wave packets passing through
  various time-dependent openings in one and two spatial dimensions.
\end{abstract}

\pacs{03.65.Nk, 03.75.-b, 42.25.Fx}

\maketitle

\section{Introduction}

Diffraction and interference of matter are among the most fascinating
and controversial aspects of quantum theory. It is not surprising that
laboratory exploration of these phenomena with subatomic, atomic, and
molecular particles has been at the heart of experimental research
since early days of quantum mechanics \cite{CSP09Optics}. As of today,
a wave-like behavior of matter has been successfully demonstrated for
a number of elementary particles, atoms, simple molecules (see
Ref.~\cite{CSP09Optics} for a comprehensive review), and, most
notably, for some heavy organic compounds including C$_{60}$
\cite{AN+99Wave}, C$_{70}$ \cite{BHU+02Matter}, and C$_{60}$F$_{48}$
\cite{HUH+03Wave}. A majority of these experiments involve sending a
mono-energetic beam of particles through a screen with openings
(apertures) such as slits, diffraction gratings, or Fresnel zone
plates. Mathematically, the problem of quantum diffraction on
stationary spatial apertures can be described by the Poisson equation
and treated by methods originally developed in the context of
diffraction and interference of light \cite{BC50mathematical,
  BW99Principles}.

A diffraction phenomenon of a different kind -- ``diffraction in
time'' -- was introduced by Moshinsky six decades ago
\cite{Mos52Diffraction} and subsequently studied by many researchers,
both experimentally \cite{SAD96Atomic} and theoretically (see
Refs.~\cite{Kle94Exact, CGM09Quantum} for reviews). The phenomenon has
to do with time evolution (more precisely, with non-uniform spreading)
of initially sharp wave fronts in quantum systems, and manifests
itself already in one dimension. Thus, in the original set-up proposed
by Moshinsky \cite{Mos52Diffraction}, a perfectly absorbing shutter is
placed in the way of a mono-energetic beam of non-relativistic quantum
particles. Then, a sudden removal of the shutter creates a ``chopped''
particle beam with a sharp wave front. As shown by Moshinsky, such a
wave front disperses non-uniformly in the course of time and, most
interestingly, develops a sequence of diffraction
fringes. Mathematically, these fringes appear to be analogous to the
ones observed in diffraction of light on the edge of a semi-infinite
plane.

There are several ways of treating quantum diffraction
theoretically. One commonly used, physically motivated method for
evaluating the wave function of a quantum particle passing through an
opening in a diffraction screen is the ``truncation'' approximation,
which is based on the composition property of quantum propagators. In
this approximation, transmission of a spatially localized wave packet
through the diffraction screen is treated as a three stage process:
(i) the wave packet is propagated freely during the time that it takes
the corresponding classical particle to reach the screen, then (ii)
the wave function is reshaped (or truncated) in accordance with the
geometry of the aperture, and, finally, (iii) the resultant wave
function is propagated freely for the remaining time interval. The
reader is referred to Refs.~\cite{FH65Quantum, Zec99Two,
  GG05Numerical, GBH+08Dispersion} for details and implementation
examples of the truncation approximation.

One drawback of the truncation approximation is that it does not
account for diffraction in time. Brukner and Zeilinger
\cite{BZ97Diffraction} proposed another, more versatile method for
solving the problem of quantum diffraction. The central assumption of
their method is that the time-dependent wave function in question
satisfies certain explicitly known, inhomogeneous time-dependent
Dirichlet boundary conditions at the surface of the diffraction
screen. More specifically, two assumptions are made: (i) the value of
the wave function at a point inside the aperture is assumed to equal
the value that the wave function would have at this point if the
diffraction screen was absent, and (ii) the wave function vanishes at
every point of the diffraction screen outside the aperture. The first
assumption is commonly referred to as the Kirchhoff approximation,
while the second corresponds to the physical assumption of {\it
  perfect reflectivity} of the screen.  The method of Brukner and
Zeilinger has since been successfully used by several researchers to
study quantum diffraction in both space and time \cite{Kal02Single,
  CM05Single, CMM07Time}.

In this paper, we present another approach to quantum diffraction in
space and time. Our method is based on the Huygens-Fresnel principle
and Kirchhoff theory of diffraction, and allows one to calculate the
time-dependent quantum propagator for the problem of particle
diffraction on spatio-temporal openings in otherwise {\it perfectly
  absorbing} screens. Our expression for the propagator is especially
adaptable for calculating diffraction patterns in situations in which
the initial wave function is given by a spatially localized wave
packet or by a superposition of several localized wave
packets. Similar to the method of Brukner and Zeilinger, our
construction provides a unified framework that treats the phenomena of
diffraction in space and diffraction in time on the same footage. The
main difference between the method of Brukner and Zeilinger and our
method is that the former is designed to treat perfectly reflecting
screens, while the latter is only applicable to perfectly absorbing
ones. Finally, our approach allows for calculation of quantum
diffraction patterns produced by openings in spatially curved screens.

The paper is organized as follows. In Sec.~\ref{sec:theor} we develop
an expansion of the propagator for a non-relativistic quantum particle
passing through a time-dependent opening in an absorbing screen. In
Sec.~\ref{sec:exam} we demonstrate utility of the expansion by
applying it to some example systems in one and two spatial
dimensions. In Sec.~\ref{sec:disc} we discuss our results and make
concluding remarks. Some technicalities are deferred to an Appendix.

\section{Huygens-Fresnel-Kirchhoff construction for quantum
  propagators}
\label{sec:theor}

In this section, we address two quantum-mechanical phenomena --
diffraction in time, pioneered by Moshinsky \cite{Mos52Diffraction,
  Kle94Exact, CGM09Quantum}, and diffraction in space, as described by
Kirchhoff theory \cite{BC50mathematical, BW99Principles}. We recast
standard descriptions of both phenomena in a way analogous to a
time-dependent formulation of the Huygens-Fresnel principle. Expressed
in this way, the two diffraction processes appear to be closely
related and can be straightforwardly unified into a single model of
diffraction in space and time.

\subsection{Notation}

In order to facilitate the clarity of the following presentation, we
begin by introducing central physical quantities and fixing notation.

The focus of this paper is on the quantum propagator $K(\vq,\vq';t)$
that describes the motion of a quantum particle in the $f$-dimensional
coordinate space by relating a particle's wave function $\Psi(\vq;t)$
at time $t$ to that at time $t=0$ through
\begin{equation}
  \Psi(\vq;t) = \int_{\mathbb{R}^f} \ud^f \! \vq' \, K(\vq,\vq';t) \Psi(\vq';0) \,.
\label{eq:prop}
\end{equation}
The propagator is the solution of the time-dependent Schr{\"o}dinger
equation
\begin{equation}
  i\hbar \frac{\partial K}{\partial t} = H_{\vq} K
\label{eq:schroedinger}
\end{equation}
with the initial condition
\begin{equation}
  \lim_{t \rightarrow 0} K(\vq,\vq';t) = \delta(\vq-\vq') \,,
\label{eq:init_cond}
\end{equation}
where $H_{\vq}$ denotes the Hamilton operator in the position
representation. (The propagator is also subject to certain absorbing
boundary conditions that we do not discuss at this point.) In the case
of a free particle of mass $m$ we have $H_{\vq} = -\frac{\hbar^2}{2m}
\boldnabla_{\vq}^2$ and $K = K_\fr (\vq-\vq';t)$ with
\begin{equation}
  K_\fr (\vq-\vq';t) = \left( \frac{m}{2\pi i\hbar t} \right)^{\frac{f}{2}}
  \exp\left( - \frac{m |\vq-\vq'|^2}{2i\hbar t} \right) \,.
\label{eq:prop_free}
\end{equation}
Hereinafter, the subscript ``$_\fr$'' indicates that the corresponding
quantity refers to the case of a free particle.

We also consider the energy-domain Green function defined as the
Laplace transform of the propagator,
\begin{align}
  G(\vq,\vq';E) &= \int_0^{\infty} \ud t \, \ue^{-st} K(\vq,\vq';t) \nonumber\\
  & \equiv \lap[K](\vq,\vq';s) \quad \mathrm{with} \quad s =
  \frac{E}{i\hbar} \,.
\label{eq:green}
\end{align}
Consequently, the propagator is obtained from the Green function by
means of the inverse Laplace transform, $K = \lap^{-1}[G]$. In the
free particle case, Eqs.~(\ref{eq:prop_free}) and (\ref{eq:green})
define the free-particle Green function $G_\fr (\vq-\vq';E) = \lap
[K_\fr] (\vq-\vq';s)$ that satisfies
\begin{equation}
  \boldnabla_{\vq}^2 G_\fr + k^2 G_\fr = -\frac{2m}{i\hbar} \delta(\vq-\vq')
  \quad \mathrm{with} \quad k^2 = \frac{2mE}{\hbar^2} \,.
\label{eq:helmholtz}
\end{equation}

\subsection{Diffraction in time}

We now address the phenomenon of diffraction in time, first considered
by Moshinsky \cite{Mos52Diffraction} and later explored by many
researchers (see Refs.~\cite{Kle94Exact, CGM09Quantum} for
reviews). In its one-dimensional formulation, the Moshinsky problem is
concerned with time evolution of a quantum particle, whose wave
function $\Psi(\xi;t)$ is localized to the semi-infinite interval
$(-\infty,x_1)$ at time $t=0$, i.e., $\Psi(\xi;0) = 0$ for $\xi >
x_1$. Over time, an absorbing wall (shutter) is switched ``on'' and
``off'' at the point $x_1$, according to a protocol defined by a
characteristic function $\chi(t)$. The latter is allowed to take
values between $0$ and $1$, with $0$ representing the case of perfect
absorption (shutter ``on'') and $1$ corresponding to perfect
transmission (shutter ``off''). The open interval $0 < \chi < 1$
represents the case of partially absorbing and partially transmitting
(but reflection-free) shutter. One is then interested in the wave
function $\Psi(\xi;t)$ of the particle to the right of the shutter,
$\xi>x_1$, at $t>0$. Moshinsky \cite{Mos52Diffraction} analyzed this
problem for a ``monochromatic'' incident wave $\Psi(\xi;0) =
\Theta(x_1-\xi) \ue^{i k \xi}$, where $k>0$ and $\Theta$ is the
Heaviside step function, and a perfectly absorbing shutter that gets
suddenly removed at an instant $t_0$, i.e., $\chi(t) =
\Theta(t-t_0)$. His analysis showed that at times $t>t_0$, the front
of the probability density wave exhibits patterns identical to those
observed in the Fresnel diffraction of light from the edge of a
semi-infinite plane, thus giving rise to the term ``diffraction in
time''.

Our objective is to devise an expression for the propagator
$K(x,x';t)$ that describes Moshinsky diffraction by a shutter
positioned at a point $x_1$, such that
\begin{equation}
  x' < x_1 < x \,,
\label{eq:x-points}
\end{equation}
and controlled (opened and closed) in accordance with the
characteristic function $\chi(t)$ of an arbitrary functional form. Our
construction relies on the Huygens-Fresnel principle, which, for the
purpose of the current problem, can be formulated as follows: The
disturbance at the point $x$ produced by a source $O'$, located at
some other point $x'$, can be viewed as produced by a fictitious
source $O_1$, located at a point $x_1$ in between $x'$ and $x$,
cf. Eq.~(\ref{eq:x-points}). The strength of the fictitious source
$O_1$ is determined by the disturbance at the point $x_1$ produced by
the original source $O'$. When a (partially) absorbing shutter is
placed at the point $x_1$, the strength of the fictitious source $O_1$
is modulated by the characteristic function $\chi(t)$ taking values
between 0 and 1. Mathematically, this can be summarized as
\begin{align}
  &K (x,x';t) = \int_0^t \ud t_1 \, u(x-x_1,x_1-x';t,t_1) \nonumber\\
  &\qquad \times K_\fr (x-x_1;t-t_1) \chi(t_1) K_\fr (x_1-x';t_1)
  \,.
\label{eq:HF_general_1D_chi}
\end{align}
where $u$, having dimensions of speed, is a yet-to-be-determined
function of the distances $x-x_1$ and $x_1-x'$ and times $t$ and
$t_1$.

The physical meaning of Eq.~(\ref{eq:HF_general_1D_chi}) is
transparent: The probability amplitude of an event in which the
particle goes from $x'$ to $x$ in time $t$ can be expressed as a sum
of probability amplitudes over all composite events in which the
particle first goes from $x'$ to an intermediate point $x_1$ in a time
$t_1$ and then reaches $x$ from $x_1$ in the remaining time
$t-t_1$. Because of their simple physical interpretation, propagator
expansions similar to Eq.~(\ref{eq:HF_general_1D_chi}) are often used
for qualitative description of quantum interference phenomena (e.g.,
see \cite{Kle94Exact} for a qualitative discussion of a two-slit
interference experiment), however the explicit functional form of $u$
is usually neither specified nor taken into account.

\subsubsection{Free particle, $\chi(t) = 1$}

Interested in determining the functional form of $u =
u(x-x_1,x_1-x';t,t_1)$, we first direct our attention to the simplest
possible scenario, in which the shutter stays open throughout the
entire time interval from 0 to $t$, i.e., $\chi(t_1) = 1$ for $0 \leq
t_1 \leq t$. In this case, the propagator in the left-hand side of
Eq.~(\ref{eq:HF_general_1D_chi}) is given by the free particle
propagator, $K = K_\fr$, requiring $u$ to satisfy
\begin{equation}
  K_\fr (x-x';t) = \! \int_0^t \! \ud t_1 \, u \, K_\fr (x-x_1;t-t_1) 
  K_\fr (x_1-x';t_1) \,.
\label{eq:HF_general_1D}
\end{equation}

It is interesting to observe that Eq.~(\ref{eq:HF_general_1D}) does
not specify the function $u$ uniquely. In fact,
Eq.~(\ref{eq:HF_general_1D}) turns out to be an identity satisfied
exactly by infinitely many different functions $u$, some examples
being (see App.~\ref{sec:u-examples})
\begin{align}
  u &= \eta \frac{x-x_1}{t-t_1} + (1-\eta) \frac{x_1-x'}{t_1} \,,
  \label{eq:u-example1}\\
  u &= \sqrt{\frac{2 i \hbar}{\pi m t}}
  \frac{\exp(-\zeta^2)}{\erfc(\zeta)} \quad \mathrm{with} \quad
  \zeta^2 = \frac{m (x-x')^2}{2 i \hbar t} \,. \label{eq:u-example2}
\end{align}
Here, $\eta$ is an arbitrary complex number, and ``$\erfc$'' denotes
the complementary error function.

\subsubsection{Moshinsky shutter, $\chi(t) = \Theta(t-t_0)$}

We now show that the functional form of $u$ can be uniquely determined
by comparing Eq.~(\ref{eq:HF_general_1D_chi}) with an exact expression
for the quantum propagator in the original Moshinsky set-up
\cite{Mos52Diffraction}, in which the absorbing shutter, located at
$x_1$, is closed until a time $t_0$ and open afterwards, i.e.,
$\chi(t_1) = \Theta(t_1-t_0)$ for $0 \leq t_0,t_1 \leq t$.

On one hand, a direct construction of the propagator
$\widetilde{K}_{\mathrm{M}} (x,x';t)$ for the original Moshinsky
problem leads to
\begin{align}
  &\widetilde{K}_{\mathrm{M}} (x,x';t) = \int_{-\infty}^{x_1} \! \ud
  x'' \, K_\fr(x-x'';t-t_0) K_\fr(x''-x';t_0) \nonumber\\ &=
  K_\fr(x-x';t) \left[ 1 - \frac{1}{2} \erfc \left( \! (x_1-x_0)
      \sqrt{\frac{m t}{2 i \hbar t_0 (t-t_0)}} \, \right) \! \right]
\label{eq:M-propagator}
\end{align}
with
\begin{equation}
  x_0 = x \frac{t_0}{t} + x' \frac{t-t_0}{t} \,.
\label{eq:x0}
\end{equation}
The first equality in Eq.~(\ref{eq:M-propagator}) combines the
composition property of quantum propagators and the fact that, at time
$t_0$, all the probability density to the right of $x_1$ has been
absorbed by the shutter. This leads to the truncation of the upper
limit in the $x''$ integral.

On the other hand, a substitution of $\chi(t_1) = \Theta(t_1-t_0)$
into Eq.~(\ref{eq:HF_general_1D_chi}) yields
\begin{align}
  &K_{\mathrm{M}} (x,x';t) = \int_{t_0}^t \ud t_1 \,
  u(x-x_1,x_1-x';t,t_1) \nonumber\\ &\qquad \times K_\fr (x-x_1;t-t_1)
  K_\fr (x_1-x';t_1) \,.
\label{eq:HF_Moshinsky}
\end{align}

Equations~(\ref{eq:M-propagator}) and (\ref{eq:HF_Moshinsky}) allow us
to uniquely determine the function form of $u$ by requiring
$\widetilde{K}_{\mathrm{M}} = K_{\mathrm{M}}$. Indeed, let us for the
moment fix the values of $x$, $x'$, and $t$, and treat the propagator
$\widetilde{K}_{\mathrm{M}}$ as a function of the shutter opening time
$t_0$ only, i.e., $\widetilde{K}_{\mathrm{M}} =
\widetilde{K}_{\mathrm{M}}(t_0)$. First, we note that $\lim_{t_0
  \rightarrow t^-} \widetilde{K}_{\mathrm{M}} = 0$. Indeed, $x_0
\rightarrow x$ as $t_0 \rightarrow t^-$, and the argument of the
complementary error function in Eq.~(\ref{eq:M-propagator}) tends to
$\ue^{i 3\pi/4} \infty$, making the value of the complementary error
function approach $2$. This limit, of course, corresponds to the
trivial case of the absorbing shutter being closed throughout the
entire time interval from 0 to $t$. Then, in view of this limit, we
rewrite the Moshinsky propagator as
\begin{equation}
  \widetilde{K}_{\mathrm{M}} (t_0) = -\int_{t_0}^t \ud t_1 \,
  \frac{\ud \widetilde{K}_{\mathrm{M}} (t_1)}{\ud t_1} \,.
\label{eq:M-propagator-int}
\end{equation}
A straightforward (but somewhat tedious) calculation yields
\begin{align}
  \frac{\ud \widetilde{K}_{\mathrm{M}} (t_1)}{\ud t_1} = &-\frac{1}{2}
  \left( \frac{x-x_1}{t-t_1} + \frac{x_1-x'}{t_1} \right) \nonumber\\
  &\times K_\fr (x-x_1;t-t_1) K_\fr (x_1-x';t_1) \,.
\label{eq:K-derivative}
\end{align}
It is now clear that the Moshinsky propagator
$\widetilde{K}_{\mathrm{M}}$, as expressed by
Eqs.~(\ref{eq:M-propagator-int}) and (\ref{eq:K-derivative}), is
equivalent to the propagator $K_{\mathrm{M}}$, given by
Eq.~(\ref{eq:HF_Moshinsky}) with
\begin{equation}
  u = \frac{1}{2} \left( \frac{x-x_1}{t-t_1} + \frac{x_1-x'}{t_1} \right) \,.
\label{eq:u}
\end{equation}
Note that Eq.~(\ref{eq:u}) is equivalent to Eq.~(\ref{eq:u-example1})
with $\eta = 1/2$. Also, Eq.~(\ref{eq:u}) provides the physical
meaning of the function $u$: The latter is a characteristic (mean)
velocity of the particle when it traverses the shutter.

\subsubsection{Arbitrary $\chi(t)$}

A substitution of Eq.~(\ref{eq:u}) into
Eq.~(\ref{eq:HF_general_1D_chi}) yields
\begin{align}
  &K (x,x';t) = \frac{1}{2} \int_0^t \ud t_1 \left(
    \frac{x-x_1}{t-t_1} + \frac{x_1-x'}{t_1}
  \right) \nonumber\\
  &\qquad \times K_\fr (x-x_1;t-t_1) \chi(t_1) K_\fr (x_1-x';t_1)
  \,.
\label{eq:HFK-1D}
\end{align}
Here, the spatial points $x$, $x'$, and $x_1$ are subject to the
condition given by Eq.~(\ref{eq:x-points}), and the characteristic
function $\chi$ is allowed to take values between 0 (perfect
absorption) and 1 (perfect transmission).

Equation~(\ref{eq:HFK-1D}) provides a formulation of the
Huygens-Fresnel principle for a one-dimensional particle in the
presence of a point-like absorbing obstacle, whose absorbing
properties change in the course of time. At this point, it is
important to emphasize that the analysis provided in this section
should not be regarded as a rigorous mathematical proof of the
propagator expansion (\ref{eq:HFK-1D}). Unavoidable difficulties in
solving the problem from the first principles stem from the lack of a
proper unambiguous definition of point-like, generally partial and
time-dependent, absorption. In our approach, however, we bypass this
issue by modeling the absorption with the help of a time-dependent
characteristic function, $\chi(t)$, and relying on the validity of the
Huygens-Fresnel construction in its most general form,
Eq.~(\ref{eq:HF_general_1D_chi}). Consequently, we remove any
arbitrariness in the Huygens-Fresnel construction by determining the
function $u$, originally unknown in Eq.~(\ref{eq:HF_general_1D_chi}),
through analyzing the case of $\chi(t)$ corresponding to the Moshinsky
shutter problem, for which point-like absorption can be defined
unambiguously.

It is interesting to note a formal similarity between
Eq.~(\ref{eq:HFK-1D}) and the well-known Lippmann-Schwinger equation
\cite{FH65Quantum, Sch81Techniques, Kle94Exact}, which in the case of
a point-like perturbation, situated at $x_1$ and described by a
spatio-temporal potential of the form $V(\xi,\tau) = \delta(\xi-x_1)
U(\tau)$, reads
\begin{align}
  &K_{\mathrm{sc}} (x,x';t) = K_\fr (x-x';t) \nonumber\\
  &-\frac{i}{\hbar} \int_0^t \ud t_1 \, K_\fr (x-x_1;t-t_1) U(t_1)
  K_{\mathrm{sc}} (x_1,x';t_1) \,.
\label{eq:LS}
\end{align}
As before, $K_\fr$ is the (free-particle) propagator in the absence of
the perturbation potential, and $K_{\mathrm{sc}}$ is the propagator
corresponding to the full scattering problem. Equation~(\ref{eq:LS})
gives rise to a multiple collision representation of the scattering
propagator, known as the Dyson series \cite{FH65Quantum,
  Sch81Techniques}.

Despite their superficial resemblance, Eqs.~(\ref{eq:HFK-1D}) and
(\ref{eq:LS}) describe very different physical processes: The
Lippmann-Schwinger equation pertains to the phenomenon of quantum
scattering, whereas the Huygens-Fresnel propagator expansion
represents quantum motion in the presence of obstacles that
(partially) absorb matter waves without deflecting them. In fact, it
is easy to show that the Lippmann-Schwinger equation can not be used
to model perfect absorption, corresponding to the trivial choice $\chi
= 0$ in Eq.~(\ref{eq:HFK-1D}). Indeed, the case of a perfectly
absorbing obstacle at $x_1$ would require $K_{\mathrm{sc}} (\xi,x';t)
= \Theta(x_1-\xi) K_\fr (\xi-x';t)$, which is clearly incompatible
with Eq.~(\ref{eq:LS}).

In order to avoid possible confusion, we emphasize that it is the
assumption of a point-like perturbation, $V(\xi,\tau) =
\delta(\xi-x_1) U(\tau)$, that does not allow the Lippmann-Schwinger
equation to properly capture the physics of absorption. On the
opposite, the Lippmann-Schwinger equation with a smooth,
complex-valued potential function $V(\xi,\tau)$, defined over an
extended spatial interval, is often the method of choice in modeling
absorbing boundaries (see Ref.~\cite{MPNE04Complex} for a
comprehensive review).

\subsubsection{Continuously opening shutter, $\chi(t) = \ue^{-\tau / t}$}

In order to demonstrate the usefulness of the Huygens-Fresnel
formulation of diffraction in time we apply the propagator expansion
(\ref{eq:HFK-1D}) to a modified Moshinsky shutter problem, in which
the absorbing shutter is initially closed, $\chi(t_1) = 0$ for $t_1
\le 0$, and then opens continuously in accordance with
\begin{equation}
  \chi(t_1) = \exp(-\tau / t_1) \quad \mathrm{for} \quad t_1 > 0 \,.
\label{eq:chi_example}
\end{equation}
Here, $\tau > 0$ determines the rate at which the shutter opens. Note
that the shutter described by Eq.~(\ref{eq:chi_example}) is completely
removed only in the limit of infinitely long times, $\lim\limits_{t_1
  \rightarrow +\infty} \chi(t_1) = 1$.

Using the Huygens-Fresnel expansion (\ref{eq:HFK-1D}), we now derive
an explicit expression for the propagator $K(x,x';t)$ connecting a
point $x'$ to the left of the shutter ($x' < x_1$) at time 0 with
another point $x$ to the right of the shutter ($x_1 < x$) at time
$t$. To this end, we rewrite Eq.~(\ref{eq:HFK-1D}), in view of
Eq.~(\ref{eq:prop_free}), as
\begin{align}
  &K (x,x';t) = -\frac{i\hbar}{2m} \left( \frac{\partial}{\partial x}
  - \frac{\partial}{\partial x'} \right) \nonumber\\ &\quad \int_0^t
  \ud t_1 K_\fr (x-x_1;t-t_1) \chi(t_1) K_\fr (x_1-x';t_1) \,.
\label{eq:HFK-1D-alternative}
\end{align}
Then, taking into account Eq.~(\ref{eq:chi_example}), we write
$\chi(t_1) K_\fr (x_1-x';t_1) = K_\fr (x_1-\tilde{x};t_1)$, where
$\tilde{x}$ satisfies
\begin{equation}
  (x_1-\tilde{x})^2 = (x_1-x')^2 + \frac{2 i \hbar \tau}{m} \,.
\label{eq:x_tilde}
\end{equation}
Solving Eq.~(\ref{eq:x_tilde}) for $\tilde{x}$ and choosing the
solution that converges to $x'$ in the limit $\tau \rightarrow 0^{+}$,
we have
\begin{equation}
  \tilde{x} = x_1 - \rho \ue^{i\theta} \,,
\label{eq:x_tilde_2}
\end{equation}
where
\begin{align}
  &\rho = \left[ (x_1-x')^4 + \left( \frac{2 \hbar \tau}{m} \right)^2
    \right]^{\frac{1}{4}} , \label{eq:rho}\\ &\theta = \frac{1}{2}
  \tan^{-1} \frac{2 \hbar \tau}{m (x_1-x')^2} \,.
\label{eq:theta}
\end{align}
Equation~(\ref{eq:HFK-1D-alternative}) can now be rewritten as
\begin{align}
  &K (x,x';t) = -\frac{i\hbar}{2m} \left( \frac{\partial}{\partial x}
  - \frac{\partial \tilde{x}}{\partial x'} \frac{\partial}{\partial
    \tilde{x}} \right) \nonumber\\ &\quad \int_0^t \ud t_1 K_\fr
  (x-x_1;t-t_1) K_\fr (x_1-\tilde{x};t_1) \,.
\label{eq:HFK-1D-alternative_2}
\end{align}
The integral over $t_1$ can now be evaluated explicitly:
\begin{align}
  \int_0^t \ud t_1 &K_\fr (x-x_1;t-t_1) K_\fr (x_1-\tilde{x};t_1)
  \nonumber\\ &= \frac{m}{2i \hbar} \, \erfc \left( \sqrt{\frac{m}{2i
      \hbar t}} (x-\tilde{x}) \right) \,.
\label{eq:int_K0_K0}
\end{align}
This identity is derived in Appendix~\ref{sec:u-examples} for the case
of real-valued $\tilde{x}$ (see Eqs.~(\ref{eq:I-int}) and
(\ref{eq:I-int3})); the derivation however can be extended to the case
of $\Re \, \tilde{x} < x_1 < x$ and $\Im \, \tilde{x} \le 0$
(cf. Eqs.~(\ref{eq:x_tilde_2}-\ref{eq:theta})). Finally, substituting
Eq.~(\ref{eq:int_K0_K0}), together with $\partial \tilde{x} / \partial
x' = (x_1-x')/(x_1-\tilde{x})$, into
Eq.~(\ref{eq:HFK-1D-alternative_2}) and taking the partial derivatives
with respect to $x$ and $\tilde{x}$, we arrive at
\begin{equation}
  K(x,x';t) = \frac{1}{2} \left( 1 + \frac{x_1-x'}{x_1-\tilde{x}}
  \right) K_\fr(x-\tilde{x};t) \,.
\label{eq:prop_chi_example}
\end{equation}

Two quick remarks are in order. First, Eq.~(\ref{eq:prop_chi_example})
guaranties that, as expected, $K(x,x';t) \rightarrow K_\fr(x-x';t)$ as
$\tau \rightarrow 0^+$. This limit corresponds to the standard
Moshinsky set-up in which the shutter stays completely open at $t >
0$. Second, it is clear from Eq.~(\ref{eq:prop_chi_example}) that the
propagator $K(x,x';t)$ is not symmetric in the coordinates $x$ and
$x'$. The absence of such symmetry is typical for quantum motion in
the presence of time-dependent obstacles (e.g., see Sec.~A.1 in
Ref.~\cite{Kle94Exact}).

\subsection{Diffraction in space}

We now address Kirchhoff theory of diffraction in space
\cite{BC50mathematical, BW99Principles}. In particular, we rewrite the
Kirchhoff's formulation in its time-dependent form, in which it can be
directly applied to diffraction of quantum wave packets.

To this end, we consider a smooth $(f-1)$-dimensional surface $S$ that
is defined as the zero set of a real-valued function $s :\,
\mathbb{R}^f \rightarrow \mathbb{R}$,
\begin{equation}
  S = \{ \vq_1 \in \mathbb{R}^f \, : \, s(\vq_1) = 0  \} \,.
\label{eq:S}
\end{equation}
The surface $S$ is assumed to partition the position space into two
disjoint regions, such that the function $s$ takes different (and
constant) signs in the two regions. We also consider two spatial
points, $\vq$ and $\vq'$, that lie on the opposite sides of the
surface. For concreteness we take $s(\vq) > 0$ and $s(\vq') < 0$. The
surface $S$ gives the location of a non-transparent, absorbing screen,
in which some transparent openings (apertures) may be ``cut
out''. These openings and, more generally, the absorbing properties of
the screen, can be described by a spatially-dependent characteristic
function $\chi(\vq_1)$ taking values between zero (perfect absorption)
and one (perfect transmission) at a point $\vq_1 \in S$. Kirchhoff
theory of diffraction \cite{BC50mathematical, BW99Principles} allows
one to express the Green function $G(\vq,\vq';E)$, connecting the
points $\vq$ and $\vq'$ at an energy $E$, as an integral along the
screen:
\begin{align}
  &G(\vq,\vq';E) = -\frac{i\hbar}{2m} \int_{\mathbb{R}^f} \ud^f \!
  \vq_1 \, \delta\big( s(\vq_1) \big) \chi(\vq_1) \nonumber\\ & \times
  \boldnabla s(\vq_1) \cdot \bigg( G_\fr (\vq-\vq_1;E)
  \boldnabla_{\!\vq_1} G_\fr (\vq_1-\vq';E) \nonumber\\ & \qquad
  \qquad -G_\fr (\vq_1-\vq';E) \boldnabla_{\!\vq_1} G_\fr
  (\vq-\vq_1;E) \bigg) \,.
\label{eq:HK_theorem}
\end{align}

It is important to emphasize that Kirchhoff method,
Eq.~(\ref{eq:HK_theorem}), is not applicable to diffraction on
apertures in transmission screens with reflecting (e.g., Dirichlet or
Neumann) boundary conditions. Instead, Kirchhoff theory is known to be
a good model for diffraction on perfectly absorbing, or ``black'',
infinitely thin obstacles \cite{NHL95Diffraction, Han95Path,
  Han01Path}. In fact, Eq.~(\ref{eq:HK_theorem}) provides an exact
solution to the time-independent diffraction problem with the boundary
conditions on the screen given by the so-called Kottler discontinuity
(see Refs.~\cite{Han95Path, Han01Path} and references within): The
probability amplitude field has a discontinuity across the screen
equal to minus the value of the free-space field at that point. A
similar condition is imposed on the normal derivative of the field.

For our purposes, it is important to construct a time-dependent
formulation of Kirchhoff diffraction. This is straightforwardly
achieved by first rewriting Eq.~(\ref{eq:HK_theorem}) as
\begin{align}
  &G(\vq,\vq';E) = -\frac{i\hbar}{2m} \int_{\mathbb{R}^f} \ud^f \!
  \vq_1 \, \delta\big( s(\vq_1) \big) \chi(\vq_1) \nonumber\\ & \times
  \boldnabla s(\vq_1) \cdot (\boldnabla_{\!\vq} - \boldnabla_{\!\vq'})
  \, G_\fr (\vq-\vq_1;E) G_\fr (\vq_1-\vq';E) \,,
\label{eq:HK_theorem-2}
\end{align}
and then performing the inverse Laplace transform from
energy-dependent Green functions to time-dependent propagators,
yielding
\begin{align}
  &K(\vq,\vq';t) = \frac{1}{2} \int_0^t \ud t_1 \int_{\mathbb{R}^f} \!
  \ud^f \! \vq_1 \, \delta\big( s(\vq_1) \big) \nonumber\\ & \quad
  \times \left( \frac{\vq-\vq_1}{t-t_1} + \frac{\vq_1-\vq'}{t_1}
  \right) \cdot \boldnabla s(\vq_1) \nonumber\\ & \quad \times K_\fr
  (\vq-\vq_1;t-t_1) \chi(\vq_1) K_\fr (\vq_1-\vq';t_1) \,.
\label{eq:Kirchhoff-time}
\end{align}

Here, an important remark is in order. The energy-domain formulation
of Kirchhoff diffraction, Eq.~(\ref{eq:HK_theorem}), only assumes the
validity of the Helmholtz equation for a field in question, and is
therefore applicable to a wide range of wave phenomena encountered,
for instance, in acoustics, optics, and non-relativistic quantum
mechanics. Equation~(\ref{eq:Kirchhoff-time}) however relies on the
particular relation, given by Eq.~(\ref{eq:green}), between the
energy-dependent Green function and time-dependent propagator, and is
restricted to non-relativistic quantum mechanics only.

Also, we note that the physical picture implied by
Eq.~(\ref{eq:Kirchhoff-time}) is that of a particle traveling freely
from $\vq'$ to a point $\vq_1$ in the aperture, and then from $\vq_1$
to $\vq$. It is therefore implicitly assumed that the aperture and
points $\vq$ and $\vq'$ are chosen in such a way that every path $\vq'
\rightarrow \vq_1 \rightarrow \vq'$ has no intersection with the
screen other than at $\vq_1$. In other words,
Eq.~(\ref{eq:Kirchhoff-time}) is only applicable to configurations in
which apertures are not ``shadowed'' by $S$ and are directly
``visible'' from the points $\vq$ and $\vq'$.

\subsection{Diffraction in space and time}

We now note a striking similarity between (i) the Huygens-Fresnel
expansion of the propagator for the problem of diffraction in time,
Eq.~(\ref{eq:HFK-1D}), and (ii) the time-dependent formulation of
Kirchhoff theory, Eq.~(\ref{eq:Kirchhoff-time}). This leads us to a
conjecture that both expressions are particular cases of a more
general propagator expansion, describing quantum diffraction on
apertures which themselves may vary in the course of time. Such
time-dependent apertures are represented by a characteristic function
$\chi$ that depends on both the position $\vq_1$ along the dividing
surface $S$ and the instant $t_1$ of the time interval $(0,t)$. As
before, $\chi = \chi(\vq_1;t_1)$ is allowed to take values between 0
(perfect absorption) and 1 (perfect transmission). In this case, we
conjecture that the propagator, connecting two points $\vq$ and $\vq'$
on the opposite sides of the screen (such that $s(\vq') < 0$ and
$s(\vq) > 0$) in time $t$, is given by
\begin{align}
  &K(\vq,\vq';t) = \frac{1}{2} \int_0^t \ud t_1 \int_{\mathbb{R}^f} \!
  \ud^f \! \vq_1 \, \delta\big( s(\vq_1) \big) \nonumber\\ & \quad
  \times \left( \frac{\vq-\vq_1}{t-t_1} + \frac{\vq_1-\vq'}{t_1}
  \right) \cdot \boldnabla s(\vq_1) \nonumber\\ & \quad \times K_\fr
  (\vq-\vq_1;t-t_1) \chi(\vq_1;t_1) K_\fr (\vq_1-\vq';t_1) \,.
\label{eq:HFK}
\end{align}
Equation~(\ref{eq:HFK}), which from now on we will refer as to
Huygens-Fresnel-Kirchhoff (HFK) construction, constitutes the main
result of the present paper.

Our arguments supporting the conjecture, given by Eq.~(\ref{eq:HFK}),
are as follows. First, the propagator expansion is in accord with the
Huygens-Fresnel principle. Second, the HFK construction correctly
captures the physics of diffraction in time: In the special case that
$\chi$ is independent of $\vq_1$ and that the surface $S$ is given by
an $(f-1)$-dimensional plane, $s(\vq_1) = {\bf n} \cdot (\vq_1-\vq_0)$
with a unit vector ${\bf n}$ and some fixed vector $\vq_0$,
Eq.~(\ref{eq:HFK}) becomes equivalent to Eq.~(\ref{eq:HFK-1D}) with
$x$, $x'$, and $x_1$ replaced, respectively, by $ {\bf n} \cdot \vq$,
$ {\bf n} \cdot \vq'$, and $ {\bf n} \cdot \vq_1$. Third, in the case
that $\chi$ is independent of $t_1$, Eq.~(\ref{eq:HFK}) is nothing but
a time-dependent formulation of Kirchhoff diffraction.

\section{Diffraction of spatially localized wave packets}
\label{sec:exam}

In this section, we apply the HFK construction, Eq.~(\ref{eq:HFK}), to
wave functions that are initially localized in position space. We
direct out attention to some example systems in one and two spatial
dimensions.

As before, we consider an absorbing screen that is defined by a
surface $S$, Eq.~(\ref{eq:S}), and a characteristic function
$\chi$. The latter, in general, is a function both of the position on
$S$ and of time. We further consider a quantum particle described at
time $t=0$ by a wave function $\Psi(\vq';0)$, which is localized in a
small neighborhood of a linear size $\sigma$ around a spatial point
$\vQ$, i.e., $\Psi(\vq';0) \simeq 0$ for $|\vq'-\vQ| \gg
\sigma$. Here, we consider the case of $\sigma$ being small compared
to the distance $L$ between the point $\vQ$ and the surface $S$. For
concreteness, we take $s(\vQ) < 0$. Then, in accordance with
Eqs.~(\ref{eq:prop}) and (\ref{eq:HFK}), the wave function $\Psi$ at a
point $\vq$, such that $s(\vq) > 0$, and time $t>0$ can be
approximately written as
\begin{align}
  &\Psi(\vq;t) \simeq \frac{1}{2} \int_0^t \ud t_1 \int_{\mathbb{R}^f}
  \!  \ud^f \! \vq_1 \, \delta\big( s(\vq_1) \big) \nonumber\\ & \quad
  \times \left( \frac{\vq-\vq_1}{t-t_1} + \frac{\vq_1-\vQ}{t_1}
  \right) \cdot \boldnabla s(\vq_1) \nonumber\\ & \quad \times
  K_\fr (\vq-\vq_1;t-t_1) \chi(\vq_1;t_1) \Psi_\fr (\vq_1;t_1) \,,
\label{eq:HFK-approx}
\end{align}
where
\begin{equation}
  \Psi_\fr (\vq_1;t_1) = \int_{\mathbb{R}^f} \ud^f \! \vq' \, K_\fr (\vq_1-\vq';t_1) \Psi(\vq';0)
\label{eq:psi-q1-t1}
\end{equation}
is the wave function of the corresponding free
particle. Equation~(\ref{eq:HFK-approx}) holds to the leading order in
the small parameter $\sigma/L$.

\subsection{One dimension}

As our first example we consider the Moshinksy problem in one
dimension ($f=1$) for a quantum particles initially described by the
Gaussian wave packet
\begin{equation}
  \Psi(x';0) = \left( \frac{1}{\pi \sigma^2} \right)^{\frac{1}{4}}
  \exp \left( \frac{i}{\hbar} P (x'-Q) - \frac{(x'-Q)^2}{2
      \sigma^2} \right) \,.
\label{eq:init_wp_1D-1}
\end{equation}
Here, $Q$ and $P$ represent, respectively, the average position and
momentum of the particle, and $\sigma$ characterizes the wave packet
dispersion in the position space. In our set-up, an absorbing shutter
is placed at a point $d$, such that $Q < d$ and $\sigma \ll L =
d-Q$. We are interested in the wave function $\Psi(x;t)$ at $x>d$ and
$t>0$.

A substitution of Eq.~(\ref{eq:init_wp_1D-1}), along with $s(x_1) =
x_1-d$ and $\chi = \chi(t_1)$, into Eqs.~(\ref{eq:HFK-approx}) and
(\ref{eq:psi-q1-t1}) yields
\begin{align}
  &\Psi(x;t) \simeq \frac{1}{2} \int_0^t \ud t_1 \left(
    \frac{x-d}{t-t_1} + \frac{d-Q}{t_1}
  \right) \nonumber\\
  &\qquad \times K_\fr (x-d;t-t_1) \chi(t_1) \Psi_\fr(d;t_1)
\label{eq:HFK-1D-approx}
\end{align}
with
\begin{align}
  \Psi_\fr &(x;t) = \left( \frac{1}{\pi \gamma_t^2 \sigma^2}
  \right)^{\frac{1}{4}} \nonumber\\ & \times\exp \left(
    \frac{i}{\hbar} \frac{P^2}{2m} t + \frac{i}{\hbar} P (x-Q_t) -
    \frac{(x-Q_t)^2}{2 \gamma_t \sigma^2} \right)
\label{eq:psi0-1D}
\end{align}
and
\begin{equation}
  Q_t = Q + \frac{P}{m} t \quad \mathrm{and} \quad
  \gamma_t = 1 + i \frac{\hbar t}{m \sigma^2} \,.
\label{eq:Q_gamma-1D}
\end{equation}
Evaluating the integral in the right-hand side of
Eq.~(\ref{eq:HFK-1D-approx}) numerically on obtains the wave function
$\Psi(x,t)$ at $x > d$ and $t > 0$.

\begin{figure}[ht]
\includegraphics[width=3.2in]{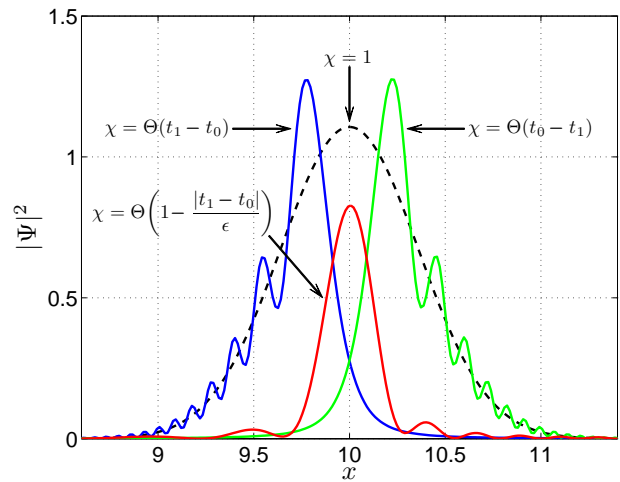}
\caption{(Color online) The probability density $|\Psi(x,t)|^2$,
  calculated in accordance with Eq.~(\ref{eq:HFK-1D-approx}) for
  various characteristic functions $\chi(t_1)$, as a function of the
  position $x$. The initial wave packet is characterized by its
  position $Q=0$, average momentum $P=200$, and dispersion
  $\sigma=0.1$. The position of the shutter is given by $d=8$, and the
  propagation time is $t=0.05$. The shutter switching time $t_0 =
  0.04$ and the time window $\epsilon = 5 \times 10^{-4}$. All
  quantities are given in the atomic units, $m = \hbar = 1$.}
\label{fig1}
\end{figure}

Figure~\ref{fig1} shows the diffraction patterns calculated in
accordance with Eq.~(\ref{eq:HFK-1D-approx}) for different shutter
protocols, $\chi(t_1)$. The initial wave packet,
Eq.~(\ref{eq:init_wp_1D-1}), is centered around $Q=0$ and has the
average momentum $P=200$ and the position dispersion
$\sigma=0.1$. Hereinafter, we use the atomic units, $m = \hbar = 1$,
in all numerical examples. The absorbing shutter is positioned at
$d=8$. Figure~\ref{fig1} shows the probability density $|\Psi(x,t)|^2$
as a function of $x$ for a fixed time, $t=0.05$. Note that, for our
choice of parameters, the position at the time $t$ of the
corresponding free classical particle is $Q_t=Q+Pt/m=10$. The figure
shows four probability density distributions for (i) $\chi(t_1) = 1$,
representing the free-particle case, (ii) $\chi(t_1) = \Theta
(t_1-t_0)$, corresponding to the perfectly absorbing shutter first
being in place and then suddenly removed at time $t_0$, (iii)
$\chi(t_1) = \Theta (t_0-t_1)$, corresponding to the shutter closed
instantaneously at $t_0$, and (iv) $\chi(t_1) = \Theta (1 -
|t_1-t_0|/\epsilon)$, representing a scenario in which the shutter is
open only during a time interval of half-width $\epsilon$ centered
around $t_0$. Here we take $t_0=0.04$ and $\epsilon = 5 \times
10^{-4}$. Note that the position at time $t_0$ of the corresponding
free classical particle is $Q+Pt_0/m = 8$ and coincides with the
position of the shutter.

Oscillations in the spatial dependence of the probability density,
seen distinctly in Fig.~\ref{fig1}, are introduced by an instantaneous
process of switching the shutter, and are, in fact, a manifestation of
the diffraction-in-time phenomenon. It is well known that these
oscillations become less pronounced and eventually disappear as one
switches the shutter ``continuously'' over longer and longer time
intervals \cite{CMM07Time, CGM09Quantum}. Here we note that the HFK
construction provides a convenient framework for analyzing the
disappearance (also known as apodization) of the diffraction pattern
for initial states that are localized in the position space. The
phenomenon of the apodization of atomic beams, which correspond to
initial states localized in the momentum space, has been previously
addressed by different methods \cite{CM05Single, CMM07Time}.

\subsection{Two dimensions}

We now address diffraction of quantum wave packets in two dimensions
on a (generally curved) shutter whose absorption properties may both
depend on position and vary in time. We consider a quantum particle
with the initial state given by
\begin{equation}
  \Psi(x,y;0) = \phi_1(x,0) \, \phi_2(y,0) \,,
\label{eq:init_wp}
\end{equation}
where
\begin{equation}
  \phi_j(\zeta,0) = \left( \frac{1}{\pi \sigma_j^2} \right)^{\frac{1}{4}}
  \exp \left( \frac{i}{\hbar} P_j (\zeta-Q_j) - \frac{(\zeta-Q_j)^2}{2
      \sigma_j^2} \right)
\label{eq:init_wp_1D}
\end{equation}
for $j = 1$ and $2$. Here, ${\bf Q} = (Q_1,Q_2)$ is the average
position and ${\bf P} = (P_1,P_2)$ average momentum of the particle,
and $\sigma_1$ and $\sigma_2$ characterize respectively the $x$- and
$y$-component of the wave packet dispersion in the position space.  As
in the one-dimensional case, the initial distance between the particle
and the shutter is assumed to be large compared to the spatial extent
of the wave packet.

Analogously to Eqs.~(\ref{eq:psi0-1D}) and (\ref{eq:Q_gamma-1D}), the
time evolution of a free particle is described by the wave function
\begin{equation}
  \Psi_\fr(x,y;t) = \phi_1(x,t) \, \phi_2(y,t) \,,
\label{eq:free_wp-2D}
\end{equation}
where
\begin{align}
  \phi_j(\zeta,t) &= \left( \frac{1}{\pi \gamma_{j,t}^2 \sigma_j^2}
  \right)^{\frac{1}{4}} \exp \left[ \frac{i}{\hbar} \frac{P_j^2}{2m} t
  \right. \nonumber\\ & \qquad \qquad \left. + \frac{i}{\hbar} P_j
    (\zeta-Q_{j,t}) - \frac{(\zeta-Q_{j,t})^2}{2 \gamma_{j,t}
      \sigma_j^2} \right]
\label{eq:free_wp_2D}
\end{align}
and
\begin{equation}
  Q_{j,t} = Q_j + \frac{P_j}{m} t \quad \mathrm{and} \quad
  \gamma_{j,t} = 1 + i \frac{\hbar t}{m \sigma_j^2} \,,
\label{eq:Q_gamma-2D}
\end{equation}
for $j=1$ and $2$. Then, using this expression for $\Psi_\fr$ in
Eq.~(\ref{eq:HFK-approx}) and evaluating numerically the
two-dimensional integral (over time and along a one-dimensional curve
accommodating the shutter), we obtain the wave function $\Psi(x,y;t)$
at $t>0$ and at a point $(x,y)$ on the side of the shutter opposite to
${\bf Q}$.

\begin{figure}[ht]
\includegraphics[width=3.2in]{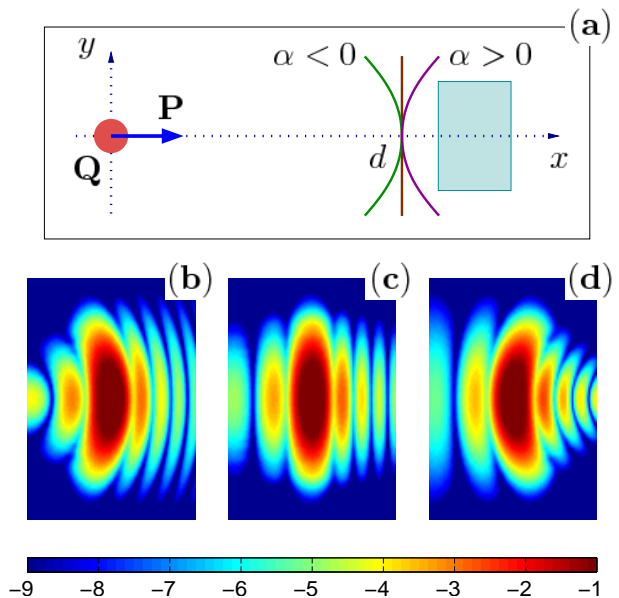}
\caption{(Color online) Diffraction in time on a curved shutter. Part
  (a) of the figure illustrates the set-up. The initial wave packet is
  characterized by its average position ${\bf Q} = (0,0)$, average
  momentum ${\bf P} = (200,0)$, and dispersion
  $\sigma_1=\sigma_2=0.1$. The shutter is positioned along the curve
  $s(x_1,y_1) = x_1 - d - \alpha y_1^2 = 0$ with $d=8$, and its time
  dependence is specified by $\chi(t_1) = \Theta
  (1-|t_1-t_0|/\epsilon)$ with $t_0 = 0.04$ and $\epsilon = 5 \times
  10^{-4}$. The propagation time is $t=0.05$. The (light blue)
  rectangle, covering $8.9 < x < 11.1$ and $-1.6 < y < 1.6$, outlines
  the spatial region, in which the probability density is
  computed. Parts (b), (c), and (d) of the figure show $\ln
  |\Psi(x,y;t)|^2$ for $\alpha = -0.3$, $0$, and $0.3$
  respectively. All quantities are given in the atomic units, $m =
  \hbar = 1$.}
\label{fig2}
\end{figure}

Figure~\ref{fig2} illustrates the effect of shutter curvature on the
diffraction pattern. The set-up is schematically presented in
Fig.~\ref{fig2}a. The initial state of the particle is given by a
Gaussian wave packet, Eqs.~(\ref{eq:init_wp}) and
(\ref{eq:init_wp_1D}), with the average position ${\bf Q} = (0,0)$,
average momentum ${\bf P} = (200,0)$, and dispersion
$\sigma_1=\sigma_2=0.1$. The wave packet is incident upon a shutter,
whose spatial geometry is defined by the curve $s(x_1,y_1) = x_1 - d -
\alpha y_1^2 = 0$ with $d=8$. Depending on the sign of the parameter
$\alpha$, the shutter is concave ($\alpha < 0$), flat ($\alpha = 0$),
or convex $(\alpha > 0)$. The time dependence of the shutter is given
by the characteristic function $\chi(\vq_1;t_1) = \Theta (1 -
|t_1-t_0|/\epsilon)$ with $t_0=0.04$ and $\epsilon = 5 \times
10^{-4}$. This corresponds to the case of a perfectly absorbing
shutter that is open only during a time interval of half-width
$\epsilon$ centered around $t_0$ (cf. solid red curve in
Fig.~\ref{fig1}). The propagation time is set to $t=0.05$, and the
wave function is investigated inside a spatial region defined by $8.9
< x < 11.1$ and $-1.6 < y < 1.6$, shown schematically as a (light
blue) rectangle in Fig.~\ref{fig2}a. Note that the position of the
corresponding classical particle at time $t$ equals ${\bf Q}_t = {\bf
  Q} + {\bf P} t/m = (10,0)$ and coincides with the center of the
rectangle. Then, the logarithm of the probability density, $\ln
|\Psi(x,y;t)|^2$, inside the rectangular region is shown for three
different values of the parameter $\alpha$: Fig.~\ref{fig2}b
corresponds to $\alpha = -0.3$ (concave shutter), Fig.~\ref{fig2}c to
$\alpha = 0$ (flat shutter), and Fig.~\ref{fig2}d to $\alpha=0.3$
(convex shutter). While the mean position of the quantum particle
coincides with that of the corresponding classical particle, the
diffraction patterns are clearly different for the three choices of
$\alpha$. Compared to the diffraction pattern produced by the flat
shutter, the concave shutter gives rise to a ``divergent'' pattern,
while the convex shutter produces a ``convergent'' diffraction
pattern.

\begin{figure}[ht]
\includegraphics[width=3.2in]{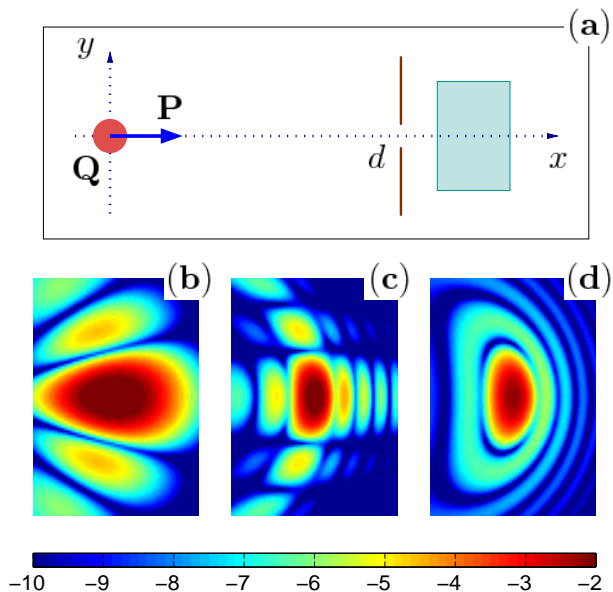}
\caption{(Color online) Diffraction on a flat absorbing screen with a
  time-dependent aperture. Part (a) of the figure illustrates the
  set-up. Parameters of the initial wave packet are the same as in
  Fig.~\ref{fig2}: ${\bf Q} = (0,0)$, ${\bf P} = (200,0)$, and
  $\sigma_1=\sigma_2=0.1$. The propagation time is $t=0.05$. The plane
  of the absorbing screen is defined by $s(x_1,y_1) = x_1-d$ with
  $d=8$. The (light blue) rectangle, covering $8.9 < x < 11.1$ and
  $-1.6 < y < 1.6$, outlines the spatial region, in which the
  probability density is computed. Parts (b), (c), and (d) of the
  figure show $\ln |\Psi(x,y;t)|^2$ for $\chi(\vq_1;t_1)$ equal,
  respectively, to $\Theta (1-|y_1|/\delta)$, $\Theta (1-|y_1|/\delta)
  \Theta (1-|t_1-t_0|/\epsilon)$, and $\big(
  1-y_1^2/\delta^2-(t_1-t_0)^2/\epsilon^2 \big)$ with $\delta = 0.05$,
  $t_0=0.04$, and $\epsilon = 5 \times 10^{-4}$. All quantities are
  given in the atomic units, $m = \hbar = 1$.}
\label{fig3}
\end{figure}

Figure~\ref{fig3} addresses diffraction on a flat screen whose
absorbing properties depend on time as well as vary in space. The
set-up, schematically illustrated in Fig.~\ref{fig3}a. As before, the
initial wave packet is described by ${\bf Q} = (0,0)$, ${\bf P} =
(200,0)$, $\sigma_1=\sigma_2=0.1$, and the propagation time is
$t=0.05$.  Here we consider a flat screen, given by $s(x_1,y_1) =
x_1-d$ with $d=8$, and study diffraction patterns for three different
characteristic functions $\chi$. As in Fig.~\ref{fig2}, we present the
logarithm of the probability density, $\ln |\Psi(x,y;t)|^2$, inside a
spatial region defined by $8.9 < x < 11.1$ and $-1.6 < y < 1.6$, and
illustrated by a (light blue) rectangle. Figure~\ref{fig3}b shows the
diffraction pattern produced by a slit in the absorbing screen and
corresponds to $\chi(\vq_1;t_1) = \Theta (1-|y_1|/\delta)$ with
$\delta = 0.05$. Figure~\ref{fig3}c represents a scenario, in which
the slit of spatial half-width $\delta$ is open only during a time
interval of half-width $\epsilon = 5 \times 10^{-4}$ centered around
time $t_0=0.04$. The corresponding characteristic function is
$\chi(\vq_1;t_1) = \Theta (1-|y_1|/\delta) \Theta
(1-|t_1-t_0|/\epsilon)$. Figure~\ref{fig3}d shows the diffraction
pattern obtained for the characteristic function $\chi(\vq_1;t_1) =
\Theta \big( 1-y_1^2/\delta^2-(t_1-t_0)^2/\epsilon^2 \big)$. In this
case, the width of the slit continuously increases from $0$ to
$2\delta$ during the time interval $t_0-\epsilon < t_1 < t_0$, and
then shrinks back to $0$ during $t_0 < t_1 < t_0+\epsilon$.

It is interesting to observe some similarity of, on one hand, the
diffraction patterns of Figs.~\ref{fig3}c and \ref{fig3}d, and, on the
other hand, patterns of Fraunhofer diffraction of light on,
respectively, rectangular and elliptic apertures in opaque
two-dimensional screens (e.g., see Figs.~8.10 and 8.12 in
Ref.~\cite{BW99Principles}). The similarity becomes even stronger when
one takes into account the space-time, $(y_1,t_1)$ representation of
the apertures leading to the diffraction patterns shown in
Figs.~\ref{fig3}c and \ref{fig3}d. Indeed, the aperture defined by the
characteristic function $\chi(\vq_1;t_1) = \Theta (1-|y_1|/\delta)
\Theta (1-|t_1-t_0|/\epsilon)$ can be represented as a rectangular
opening in the $(y_1,t_1)$ coordinates, while the aperture
corresponding to $\chi(\vq_1;t_1) = \Theta \big(
1-y_1^2/\delta^2-(t_1-t_0)^2/\epsilon^2 \big)$ has the shape of an
ellipse in the same coordinates. This observation suggests that
$f$-dimensional quantum dynamics in the presence of time-dependent
diffracting obstacles may be used for modeling diffraction in
$(f+1)$-dimensional systems with stationary, time-independent
obstacles.

\section{Conclusions}
\label{sec:disc}

In this paper we have revisited the phenomenon of diffraction in
non-relativistic quantum mechanics. More specifically, we have
considered the problem of a quantum particle passing through an
opening (or a set of openings) in a perfectly absorbing screen. Having
assumed the validity of the Huygens-Fresnel principle, we have
constructed an expansion of the quantum propagator that connects two
spatial points lying on the opposite sides of the screen. The
expansion represents the full propagator as a sum of products of two
free particle propagators, one connecting the initial point and the
other connecting the final point to a point in the opening. The
construction holds in the cases of curved (convex or concave) screens
and for openings whose shape changes in time. Consequently, our
approach allows one to analyze the quantum phenomena of diffraction in
space and diffraction in time, as well as the interplay between the
two.

In order to illustrate the method, we have used our propagator
expansion to calculate diffraction patterns for an initially localized
wave packet passing through various spatio-temporal diffraction
screens in one and two dimensions. Thus, in two dimensions, we have
investigated the effect of screen curvature on the diffraction pattern
by analyzing diffraction in time produced by absorbing parabolic
shutters and demonstrating how a convex (concave) shutter gives rise
to effective focusing (defocusing) of the wave function. We have also
studied diffraction of spatially localized wave packets passing
through holes of time-dependent size. We have shown that the shape of
a diffracted wave function is largely determined by the space-time
geometry of the hole, suggesting the use of time-dependent obstacles
in $f$-dimensional systems for modeling quantum diffraction on
stationary obstacles in $(f+1)$-dimensional systems.

The approach, developed in this paper, is applicable to quantum
diffraction on perfectly absorbing screens, and, therefore,
complements the method of Brukner and Zeilinger \cite{BZ97Diffraction}
valid for perfectly reflecting screens. In laboratory experiments,
however, diffraction screens are typically neither perfectly absorbing
nor perfectly reflecting, and the two methods have to be used in
combination. It would interesting to develop such a unified approach
and to test its predictions against results of experimental
measurements.

\acknowledgments

The author would like to thank Joshua Bodyfelt, Andre Eckardt, Orestis
Georgiou, Klaus Hornberger, Roland Ketzmerick, Klaus Richter, Roman
Schubert, Akira Shudo, Martin Sieber, and Steven Tomsovic for helpful
discussions and comments at various stages of this project.

\appendix

\section{Derivation of Eqs.~(\ref{eq:u-example1}) and
  (\ref{eq:u-example2})}
\label{sec:u-examples}

In one dimension the free-particle Green function, $G_\fr (l;E)$ with
$l>0$, is given by
\begin{align}
  G_\fr (l;E) &= \lap [K_\fr] (l;s) = \int_0^\infty \ud t \,
  \ue^{-st} K_\fr(l;t) \nonumber\\
  &= \sqrt{\frac{m}{2\pi i \hbar}} \int_0^\infty \frac{\ud
    t}{\sqrt{t}}
  \exp\left( -\frac{m l^2}{2 i \hbar t} - st \right) \nonumber\\
  &= \sqrt{\frac{m}{2i \hbar s}} \exp\left( -l \sqrt{\frac{2ms}{i
        \hbar}} \, \right) \,.
\label{app:free_green}
\end{align}
Here, $s = E/(i\hbar)$ is implied, cf. Eq.~(\ref{eq:green}), and the
reader is referred to the formula 3.471.15 of Ref.~\cite{GR07Table}
for the last integral.

We now consider the integral
\begin{equation}
  \Ical \equiv \int_0^t \ud t_1 \, K_\fr(x-x_1;t-t_1)
  K_\fr(x_1-x';t_1) \,,
\label{eq:I-int}
\end{equation}
where $x$, $x'$, and $x_1$ satisfy Eq.~(\ref{eq:x-points}).  The
Laplace transform of $\Ical$ is given by
\begin{align}
  \lap [\Ical] &= \lap [K_\fr] (x-x_1;s) \,
  \lap [K_\fr] (x_1-x';s) \nonumber\\
  &= \frac{m}{2i \hbar s} \exp\left( -(x-x') \sqrt{\frac{2ms}{i
        \hbar}} \, \right) \nonumber\\
  &= \sqrt{\frac{m}{2i \hbar s}} \, \lap [K_\fr] (x-x';s) \,.
\label{eq:I_laplace}
\end{align}
Then, using the fact that $1/\sqrt{s} = \lap [1/\sqrt{\pi t}]$ and
performing the inverse Laplace transform, we obtain
\begin{align}
  \Ical &= \sqrt{\frac{m}{2\pi i \hbar}} \int_0^t
  \frac{\ud t_1}{\sqrt{t-t_1}} \, K_\fr(x-x';t_1) \nonumber\\
  &= \frac{m}{2\pi i \hbar} \int_0^t \frac{\ud t_1}{\sqrt{t_1
      (t-t_1)}} \exp\left( -\frac{m (x-x')^2}{2 i \hbar t_1} \right) \,.
\label{eq:I-int2}
\end{align}
The integral in the last line of Eq.~(\ref{eq:I-int2}) is given by the
formula 3.471.2 of Ref.~\cite{GR07Table},
\begin{align}
  \int_0^t &\frac{\ud t_1}{\sqrt{t_1 (t-t_1)}} \exp\left(
    -\frac{\beta}{t_1} \right) \nonumber\\ &= \sqrt{\pi} \left(
    \frac{t}{\beta} \right)^{\frac{1}{4}} \exp \left(
    -\frac{\beta}{2t} \right) W_{-\frac{1}{4},\frac{1}{4}}\left(
    \frac{\beta}{t} \right) \,,
\label{eq:GR.3.471.2}
\end{align}
where $W$ stands for the Whittaker function. The latter can be written
as (see page 341 in Ref.~\cite{WW27Course}, or section 13.18(ii) of
Ref.~\cite{10Digital})
\begin{equation}
  W_{-\frac{1}{4},\frac{1}{4}} (\zeta^2) = \sqrt{\pi z} \, \exp(\zeta^2/2)
  \, \erfc(\zeta)
\label{eq:AS.13.1.33}
\end{equation}
with ``$\erfc$'' denoting the complementary error function.

Substituting Eqs.~(\ref{eq:GR.3.471.2}) and (\ref{eq:AS.13.1.33}) into
(\ref{eq:I-int2}) we obtain
\begin{equation}
  \Ical = \frac{m}{2i \hbar} \, \erfc \left( \sqrt{\frac{m}{2i \hbar t}} 
    (x-x') \right) \,.
\label{eq:I-int3}
\end{equation}
A comparison of the right-hand sides of Eqs.~(\ref{eq:I-int}) and
(\ref{eq:I-int3}) completes the proof of the identity
(\ref{eq:HF_general_1D}) with the function $u$ chosen in accordance
with Eq.~(\ref{eq:u-example2}).

In order to prove that Eq.~(\ref{eq:u-example1}) offers an alternative
choice for the function $u$ we consider
\begin{equation}
  \Ical' \equiv \left( \eta_1 \frac{\partial}{\partial x} + 
    \eta_2 \frac{\partial}{\partial x'} \right) \Ical \,,
\label{eq:Ipr-def}
\end{equation}
where $\Ical$ is given by Eq.~(\ref{eq:I-int}), and $\eta_1$ and
$\eta_2$ are two arbitrary, generally complex numbers. On one hand,
$\Ical'$ can be evaluated by a direct substitution of
Eq.~(\ref{eq:I-int}) into Eq.~(\ref{eq:Ipr-def}) which leads to
\begin{align}
  \Ical'= \frac{i m}{\hbar} \int_0^t &\ud t_1 \left( \eta_1
    \frac{x-x_1}{t-t_1} - \eta_2 \frac{x_1-x'}{t_1} \right) \nonumber\\
  & \times K_\fr(x-x_1;t-t_1) K_\fr(x_1-x';t_1) \,.
\label{eq:Ipr-2}
\end{align}
On the other hand, using Eqs.~(\ref{eq:Ipr-def}),
(\ref{eq:I_laplace}), and (\ref{app:free_green}), we have
\begin{align}
  &\lap [\Ical'] = \left( \eta_1 \frac{\partial}{\partial x}+ \eta_2
    \frac{\partial}{\partial x'} \right) \lap [\Ical] (x-x';s)
  \nonumber\\ & \qquad = \sqrt{\frac{m}{2i \hbar s}} \left( \eta_1
    \frac{\partial}{\partial x}+ \eta_2 \frac{\partial}{\partial x'}
  \right) \lap [K_\fr] (x-x';s) \nonumber\\ & \qquad = \frac{i
    m}{\hbar} (\eta_1-\eta_2) \lap [K_\fr] (x-x';s) \,,
\label{eq:Ipr-Laplace}
\end{align}
and therefore
\begin{equation}
  \Ical' = \frac{i m}{\hbar} (\eta_1-\eta_2)  K_\fr(x-x';t) \,.
\label{eq:Ipr-3}
\end{equation}
Finally, comparing Eqs.~(\ref{eq:Ipr-2}) and (\ref{eq:Ipr-3}), and
introducing a complex number $\eta = \eta_1/(\eta_1-\eta_2)$, we
arrive at the identity (\ref{eq:HF_general_1D}) with the function $u$
given by Eq.~(\ref{eq:u-example1}).


\end{document}